\documentclass[11pt]{article}
\usepackage[utf8]{inputenc}
\usepackage{booktabs}
\usepackage[dvipdfmx]{graphicx} \usepackage{bmpsize}
\usepackage{pgfplotstable}
\usepackage{float}
\usepackage{amssymb,amsmath,amsfonts, mathtools}
\usepackage{enumerate}
\usepackage{dsfont}
\usepackage[sort,compress]{cite}
\usepackage{verbatim}
\usepackage{amsfonts,euscript,amssymb,stmaryrd,braket}
\usepackage{graphics,tikz}
\usetikzlibrary{arrows,decorations.markings,patterns}
\usepackage{caption}
\usepackage{subcaption}
\usepackage{slashed}
\definecolor{hyperref}{RGB}{026,028,185}
\usepackage[bookmarks=true,colorlinks=true,linkcolor=hyperref,citecolor=hyperref,urlcolor=hyperref,bookmarksnumbered]{hyperref}
 \usepackage{booktabs}
 \usepackage{multirow}
\usepackage{tabularx}
\usepackage{longtable}
\usepackage{bbold,bbding}
\usepackage{amsthm} 
\usepackage{esint}

\newcommand{\bal}{\begin{equation}\begin{aligned}}
\newcommand{\eal}{\end{aligned} \end{equation}}

 \def\clock{{\count0=\time
           \divide\count0 60
           \ifnum\count0<10 0\fi\the\count0
           \multiply\count0 -60 \advance\count0 \time
           :\ifnum\count0<10 0\fi \the\count0
         }}
\newcommand{\timestamp}{{\small\vbox{\hbox{\tt\jobname.tex}
\hbox{\the\day/\the\month/\the\year, \clock}}}}

\newcommand{\ba}{\begin{eqnarray}}
\newcommand{\ea}{\end{eqnarray}}

\newcommand{\be}{\begin{equation}}
\newcommand{\ee}{\end{equation}}

\makeatletter
\let\old@startsection=\@startsection
\let\oldl@section=\l@section
\renewcommand{\@startsection}[6]{\old@startsection{#1}{#2}{#3}{#4}{#5}{#6\mathversion{bold}}}
\renewcommand{\l@section}[2]{\oldl@section{\mathversion{bold}#1}{#2}}
\makeatother

\numberwithin{equation}{section}
\def \x {\mathbf{x}}

\def \u {\mathbf{u}}
\def \v {\mathbf{v}}

\usepackage{color}

\newcommand{\beq}{\begin{equation}}
\newcommand{\eeq}{\end{equation}}

\newcommand{\Tr}{\textup{Tr}}

\begin{document}
\renewcommand{\thefootnote}{\arabic{footnote}}

\overfullrule=0pt
\parskip=2pt
\parindent=12pt
\headheight=0in \headsep=0in \topmargin=0in \oddsidemargin=0in

\vspace{ -3cm} \thispagestyle{empty} \vspace{-1cm}
\begin{flushright} 
\footnotesize
{HU-EP-21/40}
\end{flushright}%

\begin{center}
\vspace{1.2cm}
{\Large\bf \mathversion{bold}
{A note on improved stochastic trace estimation\\
for fermionic string fluctuations}
}
 
\author{ABC\thanks{XYZ} \and DEF\thanks{UVW} \and GHI\thanks{XYZ}}
 \vspace{0.8cm} {
%Gabriel~Bliard$^{b,c}$\footnote{{\tt gabriel.bliard@physik.hu-berlin.de}}, 
Valentina~Forini$^{a,b,}$\footnote{{\tt valentina.forini@city.ac.uk}}, Bj\"orn~Leder$^{b,}$\footnote{{\tt leder@\,physik.hu-berlin.de}},  Nils~Wauschkuhn$^{b,}$\footnote{ {\tt nils.wauschkuhn@physik.hu-berlin.de}}}
 \vskip 0.5cm   
 
 \small
{\em
%$^{a}$  II. Institut f\"ur Theoretische Physik, Universit\"at Hamburg,\\ Luruper Chaussee 149, 22761 Hamburg, Germany   
%\vskip 0.05cm
$^{a}$  Department of Mathematics, City, University of London,\\
Northampton Square, EC1V 0HB London, United Kingdom
\vskip 0.05cm
  $^{b}$  
Institut f\"ur Physik, Humboldt-Universit\"at zu Berlin and IRIS Adlershof, \\Zum Gro\ss en Windkanal 6, 12489 Berlin, Germany  
}
\normalsize

\end{center}

\vspace{0.3cm}
\begin{abstract}  
We report on the use of a stochastic trace estimator algorithm, based on mutually unbiased bases, for evaluating the trace of a matrix differential operator appearing in the context of lattice simulations for the discretized superstring worldsheet. 
A study of the variance,  in a setup which is slightly modified with respect to the original one,  confirms advantages with respect to more traditional methods like the Gaussian estimator.

% RESUB , first presented in~\cite{Forini:2017mpu},

% the discretized worldsheet of Type IIB strings in the Gubser-Klebanov-Polyakov background in a new setup, which eliminates a complex phase previously detected in the fermionic determinant. A sign ambiguity remains, which a study of the fermionic spectrum shows to be related to Yukawa-like terms, including those present in the original Lagrangian before the linearization standard in a lattice QFT approach.  Monte Carlo simulations are performed in a large  region of the parameter space, where the sign problem starts becoming severe and instabilities appear due to the zero eigenvalues of the fermionic operator.  To face these problems, simulations are conducted using the absolute value of a fermionic Pfaffian obtained introducing a small twisted-mass term, acting as an infrared regulator,  into the action.
%modified with an infrared regulator. 
%The sign of the Pfaffian and the low modes of the quadratic fermionic operator are then taken into account by a reweighting procedure of which we discuss the impact on the measurement of the observables. In this setup we study bosonic and fermionic correlators and observe a divergence in the latter, which we argue - also via a one-loop analysis in lattice perturbation theory -  to originate from the U(1)-breaking of our Wilson-like discretization for the fermionic sector.  

\end{abstract}

%%%%%%%%%%%%%%%%%%%%%%%%%%%%%%%%%%%%%%%%%%%%%
%%%%%%%%%%%%%%%%%%%%%%%%%%%%%%%%%%%%%%%%%%%%%
%\tableofcontents
  
\section{Discussion}

The stochastic estimation of traces, a way to approximate the trace of a matrix (not explicitly known) using only matrix vector products,  is a problem of large interest in physics and applied mathematics in general. A stochastic trace estimator chooses 
a random,  $n$-dimensional probe (complex) vector $\x$ so that the expectation of $\x^\dagger \,A\,\x$ for a given matrix $A$ equates the trace of $A$. Among the many possible approaches to sampling, the proposal of a mutually unbiased bases (MUBs) estimator~\cite{fitz} is one which requires that the absolute value of the overlap between pairs of vectors drawn from  bases corresponding to different measurements (here one deals with orthonormal bases, from whose elements $x$ is chosen uniformly at random)  is constant.  This method has not been so far used in the research area of lattice field theory, and it appears an interesting exercise to see whether also in this context the comparison to traditional methods, such for example the Gaussian estimator, reveals possible advantages in terms of variance.   Other ways to analyse an estimator consider theoretical bounds on the number of samples required to achieve an $\epsilon$-approximation of the trace, as well as the number of bits required to generate the complex vector $\x$~\cite{avron}.

In this short note, we report on the use of the MUBs estimator algorithm to evaluate the trace of a matrix differential operator which appears in the lattice simulations, via Rational Hybrid Monte Carlo algorithm, for the discretized string worldsheet model of~\cite{POS2015,Bianchi:2016cyv,Forini:2017mpu,Forini:2017ene,Bianchi:2019ygz}. 
The original motivation of this exercise was performing  numerical simulations in order to measure the main observable investigated in~\cite{Roiban,POS2015,Bianchi:2016cyv,Forini:2017mpu,Bianchi:2019ygz}, the cusp anomalous dimension  of $\mathcal{N}=4$ super Yang-Mills. This can be done in principle using an alternative definition  which involves a trace (see Appendix~\ref{app:string}), and the new set of measurements was initially an attempt to shed more light on some aspects of the difficulties encountered in~\cite{Bianchi:2019ygz}, where the Wilson-like fermionic discretization breaks part of  the global symmetry of the model. 
In a parallel work~\cite{Bliard:2022kne}, however, we present a novel discretisation which preserves all global symmetries. For it, a preliminary lattice perturbation theory analysis suggests that, in order to remove UV divergences at one loop, it is necessary to introduce two extra parameters in the action, which need to be either ﬁne-tuned at tree level or renormalized at one-loop.   It is therefore rather in this new setup that - once enough evidence for the existence of the continuum limit at least perturbatively is established~\footnote{As discussed in~\cite{Bliard:2022kne},  where one-point functions and one bosonic correlator of the worldsheet excitations were considered, a \emph{complete} one-loop analysis of the divergences of general n-point functions may shed light on this issue. Clearly, only an all-order argument would be needed to establish that this discretized model is a non-perturbative deﬁnition of the AdS$_5\times S^5$ string in null-cusp background.} - one should proceed with measurements via Monte Carlo simulations. Here, we still find useful to report an interesting byproduct of the study in the setup of~\cite{Bianchi:2019ygz}, namely a concise analysis -- which considers only the  variance --  of the MUBs estimator in a high-energy-physics research area.  

%Before proceeding with numerical simulations in that setup, to establish the existence of the continuum limit at least at perturbative level a complete study of n-point functions at one loop would be needed, 
%show that that even when working with a discretised action which preserves all global symmetries, working with the parameter space of the original lagrangian is not enough to ensure a continuum limit at one loop in lattice perturbation theory.  This prevents -- at least until some strong evidence is given for perturbative renormalizability of the discretised gauge-fixed sigma-model of interest -- from considering numerical simulations in general a viable study tool for this setup.  

The upshot is that, when compared to the Gaussian trace estimator, the MUBs estimator shows a significant reduction of the variance\footnote{Our definition of variance here coincides with what in~\cite{fitz} is called ``single shot sampling variance'', namely $V/N_s$ with $N_s=1$.}, 
implying that for a fixed sample variance ($V/N_s$) a smaller number of samples $N_s$ is needed. This is a clear advantage in terms of simulation costs, although in the present case the most important contribution to such costs is related to a matrix fermion inversion.  In our application an important improvement is obtained using the MUBs algorithm excluding the standard (canonical) basis, what we call MUBw (w for without) below. 
%\colb{[Maybe later] The improvement is observed for the matrices considered here, and  in general for $n\times n$-matrices $A$ if  $\sum_{i=1}^n A^2_{ii}-\frac{1}{n+1}\Tr(A^2)>\frac{1}{n+1}\Tr(A)^2$.}  
The latter happens to have the same variance as the Hutchinson algorithm~\cite{hutch}, %(used with complex numbers for symmetric matrices).  
but still outperforms  it in terms of amount of random bits required to create the probe vectors $x$.  The fact that the number of bits required grows only logarithmically, as opposed to the linear growth of the Hutchinson algorithm, is a general feature of the MUBs algorithm (with or without the standard basis, and independently on the application)~\cite{fitz}.
 %Such a crucial decrease in the amount of randomness required is in fact a general feature of the MUB algorithm~\bcite{fitz}, independently on the exclusion of the standard basis and on the application. 
%It would be interesting to test the algorithm on some  - e.g. like stochastic estimation with $Z_2$ noise~\cite{StochEstZ2Noise} and to estimate the trace appearing in the action-derivative of \cite[eq. (4.13)]{HeavySeaQuarks} in lattice quantum chromodynamics.

 As we discuss below, an improvement of the method would be the treatment or removal of the tails in the distribution without introducing large bias to the estimator. A direct comparison with the Hutchinson algorithm (stochastic estimation with $Z_2$ noise~\cite{Dong:1993pk}) would be of great interest from a theoretical point of view (why are the variances so similar or even equal?) and from an application point of view in, e.g., lattice quantum chromodynamics.
%\emph{TO CHANGE. An analysis of the histogram and data of the estimations with all vectors of all MUB bases except the standard basis, we found a disadvantage of the MUB methods: 
%The disadvantage of the MUB algorithms are the outliers lying far away from the mean value: In a tested case the excess kurtosis had the order $10^5$. To optimize the variance of the algorithm for a given sort of matrices, it could be researched if excluding further bases reduces the variance and the excess kurtosis. \\}

%OUTLOOK
%The exaple we consider
%It would be interesting to motivate the improvement that we observe via an analysis of possible cancellations of higher moments of the distribution on which we take expectation values . \colb{but don't we have soe statement about kurtosis??}

%maybe the improvement is due to the fact that the higher moments of the distribution on which we are taking expectation values we succeed in cancelling them? Because tipically  they would give noise.

%A: Higher moments contribute to the fluctuations (noise) in an actual computation and they might improve or make it worse. 

%To understand whether there is cancellation one would have to write down the formula, and see whether the contribution to the error from the second moment is cancelled even partly by the contribution from the fourth moment, for some range of parameters.

%But we did not compute the 4th moment

%The only thing that we have is the leading contribution to the error, which is the second moment (or variance), and we cannot say more. 
%
%
%
In the following, after  presenting the matrix differential operator of interest, we proceed recalling few facts about the MUBs estimator (following~\cite{fitz}) and apply it to the case of the fermionic operator describing string fluctuations above the null cusp solution. In Appendix~\ref{app:string} we place in its context the matrix differential operator of which, at finite lattice spacing, we evaluate the trace. 

\section{Setup and trace estimator}

We are interested in evaluating the trace of a matrix differential operator, which below we call $\mathbb{O_F}$, whose emergence in the context of string theory and the AdS/CFT correspondence is discussed in Appendix~\ref{app:string} (see also a summary in~\cite{Roiban,Bianchi:2016cyv,Bianchi:2019ygz}).  There we explain that a possible definition of the cusp anomaly of $ \mathcal{N}=4$ super Yang-Mills (alternative to the one used in~\cite{Bianchi:2016cyv, Bianchi:2019ygz,toappear}) uses the logarithmic derivative of the string partition function with respect to the massive parameter $m$, which together with the string tension $g$ (sigma-model effective coupling)  is one of the two bare parameters of the model. This implies -- see \eqref{eq:TraceSimplification} -- that an interesting quantity to measure via numerical Monte Carlo simulations in the discretized model is  
\be\label{trace}
\Tr\, \mathbb{O_F}\equiv \langle\Tr \left(O_{F}(0)O_{F}(m)^{-1} \right)\rangle\,.
\ee
In the continuum theory, $O_F(m)$ -- for which emphasizing the dependence on $m$ is just a matter of convenience -- is a $16\times 16$ matrix differential operator which is conveniently written in $4\times4$ blocks
\be
O_F(m)=
 \begin{pmatrix}  0 & i \partial_t & -i \rho_M(\partial_s+\frac{m}{2})\frac{z^M}{z^3} & 0  \\
                    i \partial_t & 0 & 0 & -i\rho^{\dagger}_M(\partial_s+\frac{m}{2})\frac{z^M}{z^3}  \\
                    i \rho_M(\partial_s-\frac{m}{2})\frac{z^M}{z^3} & 0 & 2\rho_M(\partial_s x-\frac{m x}{2})\frac{z^M}{z^4} & i \partial_t - A^{\text T}\\
                    0&i\rho^{\dagger}_M(\partial_s-\frac{m}{2})\frac{z^M}{z^3}&i\partial_t+A&-2\rho^{\dagger}_M(\partial_s x^* -\frac{m x^*}{2})\frac{z^M}{z^4}
                 \end{pmatrix},
\label{OF}
\ee
with
\be\label{A}
 A=-\frac{\sqrt{6}}{z}\phi 1\!\!1 +\frac{1}{z} \tilde{\phi}+\frac{1}{z^3}\rho^*_N \tilde{\phi}^{\text T}\rho_L z^N z^L +i \frac{z_N}{z^2} \rho^{MN} \partial_t z^M.
\ee
Above, the $4\times4$ matrices $(\rho^M)_{ij}$  are the off-diagonal blocks of the Dirac $\gamma^M$ matrices of $SO(6)$, and $\rho^{MN}$ are the generators of $SO(6)$ with
$(\rho^{MN})^i_j:=\frac{1}{2}\left[\left(\rho_M\right)^{il} \left(\rho_N\right)_{lj}-\left(\rho_N\right)^{il} \left(\rho_M\right)_{lj} \right]$ and $(\rho_M)^{ij}=((\rho_M)_{ji})^*$. For an explicit representation of them, see for example Appendix A of~\cite{Bianchi:2019ygz}. 
In  the entries of the matrix above bosonic fields and their derivatives appear, function of two (worldsheet) variables $(t,s)$. They are the string (target space) coordinates --  $x(t,s)$, $x^*(t,s)$ (two complex bosonic $AdS_5$ coordinate fields orthogonal to the surface of the relevant classical solution) and  $z^M$, with $\sqrt{z_M z^M}=z$ (further bosonic coordinates of $AdS_5 \times S^5$ in Poincar\'e patch) -- together with  certain auxiliary fields $\phi, \phi_I$, $I=1,...,16$, arising from a Hubbard-Stratonovich transformation~\cite{Bianchi:2019ygz}.   
The  matrix $\tilde{\phi}$ in~\eqref{A} consists of the 16 real (auxiliary) fields $\phi_I$,
\be
\tilde{\phi}=\frac{1}{\sqrt{2}}
 \begin{pmatrix}  \sqrt{2}\phi_{13} & \phi_1+i \phi_2 & \phi_3+i\phi_4 & \phi_5+i\phi_6  \\
                    \phi_1-i\phi_2 & \sqrt{2}\phi_{14}  & \phi_7+i\phi_8 & \phi_9+i\phi_{10} \\
                    \phi_3-i\phi_4& \phi_7-i\phi_8 & \sqrt{2}\phi_{15}  & \phi_{11}+i\phi_{12}\\
                   \phi_5-i\phi_6&\phi_9-i\phi_{10}&\phi_{11}-i\phi_{12}&\sqrt{2}\phi_{16} 
\end{pmatrix}.
\ee
In the discretized setup of~\cite{Bianchi:2019ygz}, to avoid fermion doublers a Wilson-like operator is added to $O_F$ (in Fourier space, this is $W_\pm$ in~\eqref{OFWilson}-\eqref{Wilsonshiftgen}). Then, expectation values of the observables of interest  can be obtained by Monte Carlo simulations~\footnote{Due to $W_\pm$ breaking part of the global symmetry of the model, divergences occur, see discussion in~\cite{Bianchi:2019ygz}.} for the square lattice $N_\text{t}\times 2 N_\text{sp}$, where $N_\text{t}\equiv N_\text{sp}$ the number of lattice points in time and space direction, and lattice spacing $a$ (for simplicity, below we denote the number of lattice points simply by $N$, the physical length is then $L=N\,a$, for the time length we have been using $T=2L$). Here we use a Rational Hybrid Monte Carlo (RHMC) algorithm~\cite{RHMC1,RHMC2}, generating a sample of (bosonic) field configurations on which the discretized operator $O_F(m)$ in~\eqref{OFWilson} can be evaluated.  In addition to this ``outer'' RHMC, for each field configuration an ``inner'' Monte Carlo algorithm is needed to estimate the trace~\eqref{trace} of the implicit matrix $O_F(0)O_F(m)^{-1}$.  This is most efficiently done~--~once $O_F(0)O_F(m)^{-1}$ is evaluated, for which we employ a multi-shift conjugate gradient solver whose implementation is inspired from the  \texttt{openQCD} package~\footnote{https://luscher.web.cern.ch/luscher/openQCD/. See discussion in~\cite{Bianchi:2019ygz}.}  -- by using a stochastic trace estimator based on matrix-vector products. 

\bigskip
 
 As mentioned above, to evaluate the trace of a square matrix $A$ of size $n$ stochastically one chooses an $n$-dimensional probe complex vector $\x$ uniformly at random from a given orthonormal basis $B$, 
so that the trace is exactly given by  $\text{Tr}A=\sum_{ \x\in B} \x^ \dagger A \,\x$.  The sampling policy distinguish different trace estimators, and here we concentrate on the one in which the basis $B$ is also chosen uniformly at random from a set of $b$ mutually unbiased bases (MUBs) $\mathcal{B}=\{B_1,\ldots,B_b\}$, where $b$ is taken to be the maximum number of mutually unbiased bases for a complex vector space of dimension $n$~\cite{fitz}. A set of orthonormal bases $\{B_1,\ldots,B_b\}$ is said to be of mutually unbiased bases if for all choices of bases $i$ and $j$, $i\neq j$ within it (so, bases corresponding to different measurements)  the absolute value of the overlap between every $\u \in B_i$ and every $\v \in B_j$ is constant. In particular, the requirement is $|\u^\dagger \v| = \frac{1}{\sqrt{n}}$, where $n$ is the dimension of the space. In the case of complex vector spaces the number of mutually unbiased bases is $n+1$, in the case in which $n$ is either a prime or a prime raised to some integer power~\cite{klapp}.
In the case in which $n$ is not falling into this category -- which is our case -- the matrix can be always filled up with zeros until the dimension of a prime is reached. 

Obviously the presence of these zeros leads to outliers in the distribution of the trace estimation and therefore to a larger variance.  
%Obviously the presence of these zeros will result in zero estimates for the trace. Especially in the case of few estimations, this leads  to an estimated sample variance which is large and an average of the estimations which is far away from the exact trace. 
To prevent this effect, one can exclude from $\mathcal{B}$ the standard basis (constructed out of the vectors which build the identity matrix, we call it $\mathcal{B}_s$ below), working with a slightly modified algorithm which we denote with MUBw. As we see below, the exact variance for the MUBw estimator is then modified, and equates the exact variance for the Hutchinson estimator. 
While this formal change has little impact in the application, removing the outliers leads effectively to a big improvement for the sample variance.     In the next section, we  discuss the presence of other outliers that impact the sample variance.

We will compare the MUBw algorithm  with the Gaussian trace estimator. For it, the estimation of the trace of $A$ is $\x^{\dagger}\, A\, \x$,  the elements of the vector $\x$ are independent and identically distributed random numbers of a Gaussian distribution with unit variance and zero mean.

\bigskip
 
 We refer  the reader to~\cite{fitz} for a  thorough description of the MUBs trace estimator (and a condensed account of estimators such as the fixed basis estimator, Hutchinson, Gaussian) and the algebraic steps to derive the theoretical definition of its variance. Here we just recall some definition and result. Given a random variable $X$ and its expectation value $E(X)$,  
 the variance for a single trial of $X$ is
 \be\label{variance_def}
\text{ Var}(X)=E(X^2)-E(X)^2\,.
 \ee 
In our case the random variable $X$ is  $X=\x^\dagger A\x$, where $\x$ is chosen according to the policy of the given trace estimator. In the case of MUB, one has  
 \be
 E(X)=\frac{1}{n\,b}\sum_{B\in \mathcal{B}}\sum_{{\x}\in B}\,\x^\dagger A\,\x=\frac{1}{n} \text{Tr}A 
 \ee
% \,X=x^\dagger A\,x
so that the second term in~\eqref{variance_def} equates $\frac{(\text{Tr}A)^2}{n^2}$. 
The first term in~\eqref{variance_def} reads formally as follows in terms of the Kronecker product ($\otimes$) of matrices
\begin{equation}\label{exsqmub}
E(X^2) = \frac{1}{nb}\sum_{B\in \mathbb{B}} \sum_{\x \in B} \left(\x^\dagger A \x\right)^2 = \frac{2}{nb} \text{Tr}\left(P  A^{\otimes 2} \right)\,,\qquad\qquad P=\frac{1}{2}\sum_{B\in \mathbb{B}} \sum_{\x \in B}\left(\x \x^\dagger\right)^{\otimes 2}\,.
\end{equation}
Some algebra~\footnote{ $P$ in~\eqref{exsqmub} can be proven to  be a positive semi-definite projector. Using this, the spectral decomposition for $A$ and the fact that $n$ is a prime number or a prime raised to some integer power one arrives to the final formula~\eqref{varianceMUB}. See~\cite{fitz} for the full derivation. } is then needed~\cite{fitz} to arrive to the simple final form for the variance of the MUBs estimator
\be\label{varianceMUB}
V_\text{MUBs}=\text{E}(X^2)-\text{E}(X)^2=\frac{n}{n+1}\Tr(A^2)-\frac{1}{n+1} (\Tr A)^2\,.
\ee
To define the MUBs estimator \emph{without} the standard basis $\mathcal{B}_s$, MUBw, one starts from the direct subtractions in the definitions of the expectation values
 \begin{eqnarray}
 E(X)_\text{MUBw}&=&\frac{1}{n\,b}\sum_{B\in \mathcal{B}}\sum x^\dagger\,A\,x -\frac{1}{n}\sum_{e_i\in \mathcal{B}_s}\,e_i^\dagger\,A\, e_i\,,\\
 E(X^2)_\text{MUBw}&=&\frac{1}{n\,b}\sum_{B\in \mathcal{B}}\sum (x^\dagger\,A\,x)^2-\frac{1}{n}\sum_{e_i\in \mathcal{B}_s}\,(e_i^\dagger\,A\, e_i)^2\,,
 \end{eqnarray}
 where $e_i$ are the canonical vectors (made of zeros and a single unit entry) building the identity matrix.  Proceeding then with similar algebraic steps as in~\cite{fitz}, one obtains for the theoretical variance the  formula
\be
V_\text{MUBw}=\text{E}(X^2)_\text{MUBw}-\big(\text{E}(X)_\text{MUBw}\big)^2=\text{Tr}(A^2)-\sum_i A^2_{ii}\,.
\ee
Notice that this coincides with the variance for the often used Hutchinson estimator~\cite{avron,hutch}. The latter, however, is dramatically worse in terms of random bits required, which are in the order of $n$~\cite{avron,hutch}. In the case of the MUB estimator~\cite{fitz},  at least $[\log_2 (n)]$ random bits are required to allow the possibility of choosing every element $\x$ of a fixed basis $B$, and  $[\log_2(b)]$ are necessary to choose B (recall that $b=n+1$, when $n$ is a prime or a prime raised to some integer power). 
The Gaussian estimator with complex random vectors for a symmetric matrix has theoretical variance $V_{Gauss}=\text{E}(X^2)-\text{E}(X)^2=\Tr(A^2)$ (in the case of real random vectors this is $2\Tr(A^2)$~\cite{avron}).
 %; the complex case complex random vectors the variance behaves like having two independent real estimations because real and imaginary part are independent). 
 %$X$ is the estimation of the trace with the respective algorithm.  

Summarizing, the analytic expectations for the variance and the random requirements for the stochastic trace estimation algorithms considered here are as follows%~\colb{Table di fitsimons}\\

 \vspace{0.2cm}

\begin{tabular}{lcc}
\underline{Trace estimator}&\underline{Variance $V$}&\underline{Random bits required}\\%&\underline{remarks}\\
Gaussian&$\text{Tr}(A^2)$&$\mathcal{O}(n)$\\ %&\\
Hutchinson&$\text{Tr}(A^2)-\sum_i A^2_{ii}$&$\mathcal{O}(n)$\\%&\\
MUBw&$\text{Tr}(A^2)-\sum_i A^2_{ii}$&$\mathcal{O}(\log(n))$\\%&outliers lying far away\\
MUBs&$\frac{n}{n+1}\text{Tr}(A^2)-\frac{1}{n+1}(\text{Tr}(A))^2$&$\mathcal{O}(\log(n))$\\%&outliers lying far away\\
\end{tabular}

\section{Results}

\begin{figure}[h]
\begin{center}
\includegraphics[width=15cm]{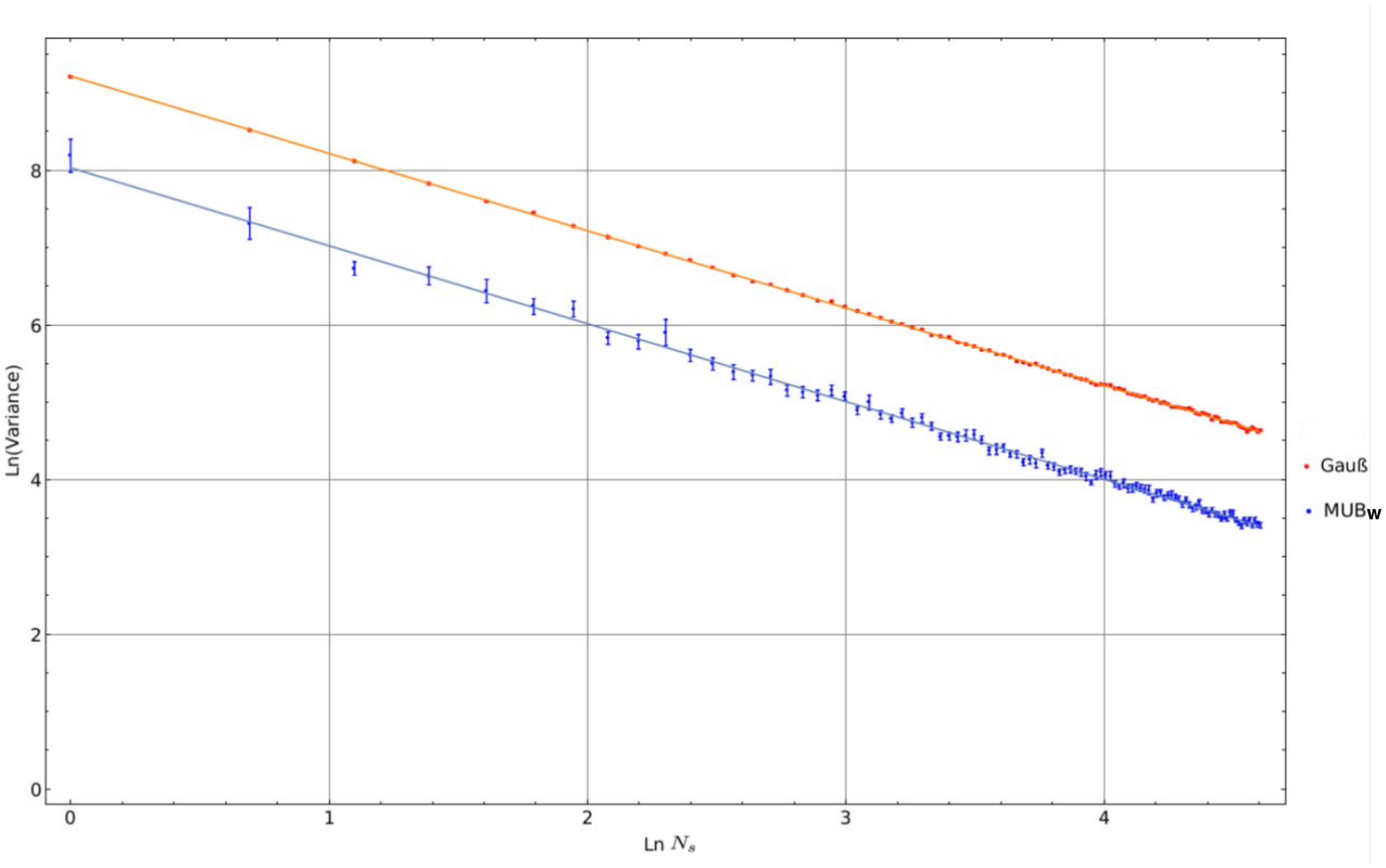}
\vspace{-1.5cm}
\caption{Logarithm of the variance for the averages of $N_s$ estimations of the trace of $O_F^\dagger O_F$ using random (bosonic) field configurations, with linear fits for MUBw (blue) and Gaussian (red) estimators. 
Error bars are obtained by repeating the variance estimations $100$ times for each value of $N_s$.}
\label{fig:variance}
\end{center}
\end{figure}

To first test the algorithm we have applied it to the matrix $O_F^\dagger\,O_F$~\footnote{ 
The matrix $O_F^\dagger\,O_F$ is  symmetric positive semi-definite, the case considered in~\cite{fitz}. While this feature is not required for the calculation of formula~\eqref{varianceMUB}, it is however necessary for working out the theoretical bounds of~\cite{avron} or for discussing the better worst case performance, see discussion in~\cite{fitz}. Our object of interest is the matrix $O_F(0)O_F(m)^{-1}$, which is not symmetric positive semi-definite. In our study we therefore just monitor the variance  and how it is included in the whole stochastic error analysis.}. 
%~\footnote{The feature of being symmetric and positive definite is relevant for the interpretation....}. 
While the run-time for one estimation was nearly the same for both Gaussian and MUBs estimators, a clear difference is found in the sample variance for the number (size) $N_s$ of samples.  
The latter is connected to the analytic expectation $V$ in the table above by $V/N_s$. Such functional form (with standard corrections of order $1/N_s^2$) is clearly supported by the linear fits, for the logarithm of the variance,  in Fig.\ref{fig:variance}.  
 We read off from the graph an advantage for the variance by a  factor of about $3$.
\vspace{-1.5cm}
\newpage
An even larger improvement in variance can be observed for the case of $O_F(0)O_F(m)^{-1}$, which is our main application. 
%~\footnote{
%%Notice that $O_F O_F^\dagger$ is a symmetric and positive semi-definite matrix, a feature which was exploited in~\cite{fitz} to prove an improvement of the MUBs estimator in terms of worse case performance. 
%}. 
%
% For the observed cases, the improvement of the variance of the MUB algorithm in contrast to the Gaussian algorithm is larger with $O_F(0) O_F(m)^{-1}$ compared with the case of symmetric positive semidefinite matrices $O_{F,Lat}O^{\dagger}_{F,Lat}$:
For a $(N_\text{t}=16) \times (N_\text{sp}=8)$-lattice and $900$ estimations~\footnote{$30$ variance estimations, repeated $30$ times to estimate the error.},  we found the following results:
\begin{center}
\begin{tabular}{ccc}
\textbf{Method}&\textbf{Variance}&\textbf{Trace}\\
\hline
MUB&41$\pm$20&1865.5\\
MUBw&25$\pm$10&1865.5\\
Gaussian&1750$\pm$89&1863\\
\textit{Exact trace}&-&\textit{1865.5}\\
\end{tabular}\\
\normalsize
\hspace{2cm}\\
\end{center}
Table 1: Variance and trace estimation of $O_F(0)O_F(m)^{-1}$ with the MUB method, the MUB method without the standard basis (MUBw) and the Gaussian method using random (bosonic) field configurations and values $g=5, r=1, L\,m=8$ for the parameters appearing in $O_F$, see Appendix~\ref{app:string}.  
 %The exact trace could be calculated, for this case of a small system, using an explicit implementation of $O_F$ to compute its inverse. 
\bigskip
%
%
%
%\FloatBarrier
%\footnotetext{the estimations outside the plotted range can't be seen, the standard basis was not used}

\bigskip

\begin{figure}
  \begin{minipage}[c]{0.6\textwidth}
    \includegraphics[width=\textwidth]{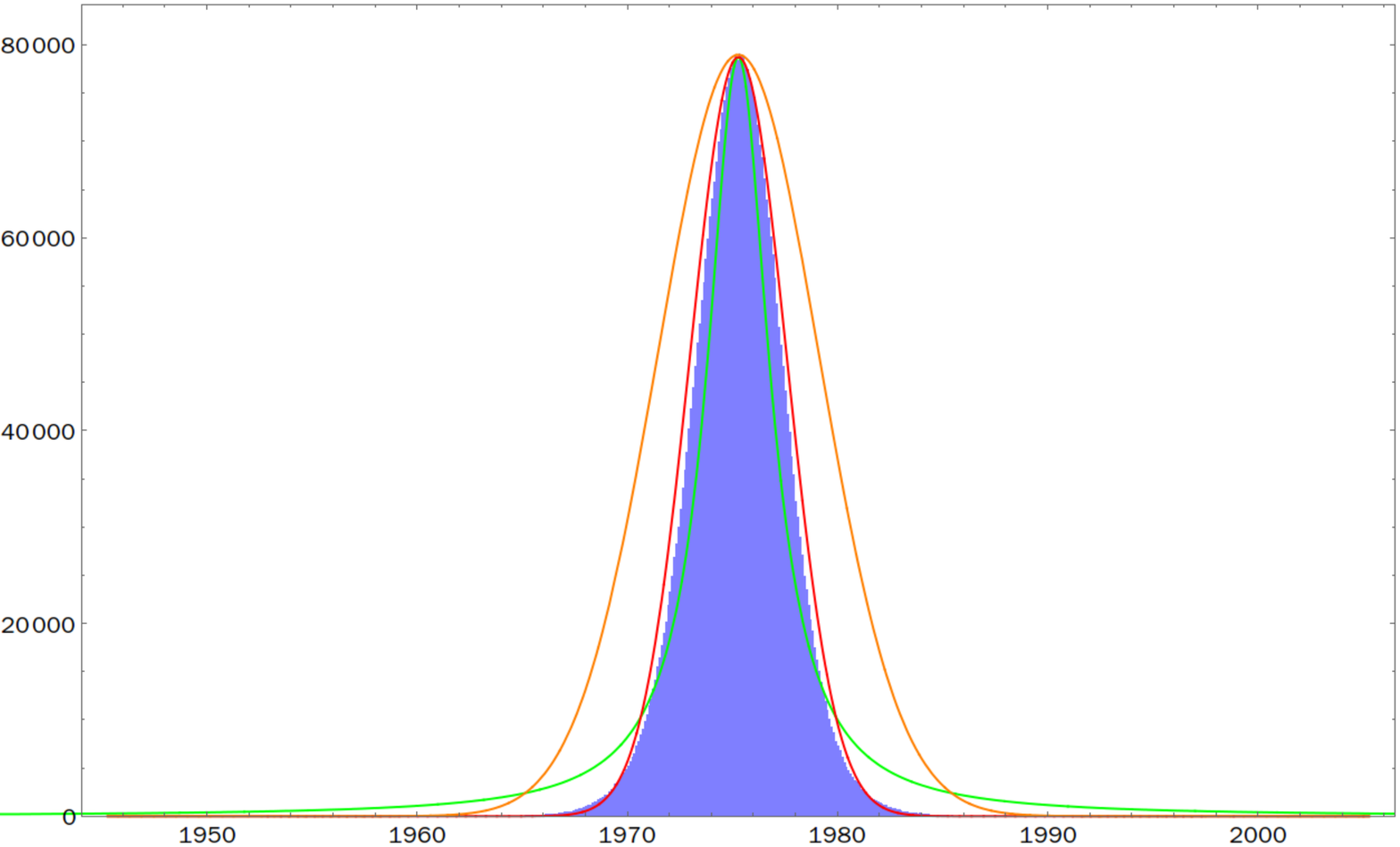}
  \end{minipage} \hfill
  \begin{minipage}[c]{0.35\textwidth}
    \caption{Histogram of %``all'' 
    estimations of $\text{Tr}\mathbb{O_F}$ with the MUBw algorithm for a realistic configuration from a MC simulation with parameters $N=8$,~$g=50$,~$Lm=4$. The red and green lines are best fit Gaussian and Cauchy distributions with exact trace as mean and height fixed. The orange line is a Gaussian distribution with the exact variance of the MUBw method. 
    \label{fig:histmub}
    }
  \end{minipage}
  
\vspace{0.00mm}

  \begin{minipage}[c]{0.6\textwidth}
    \includegraphics[width=\textwidth]{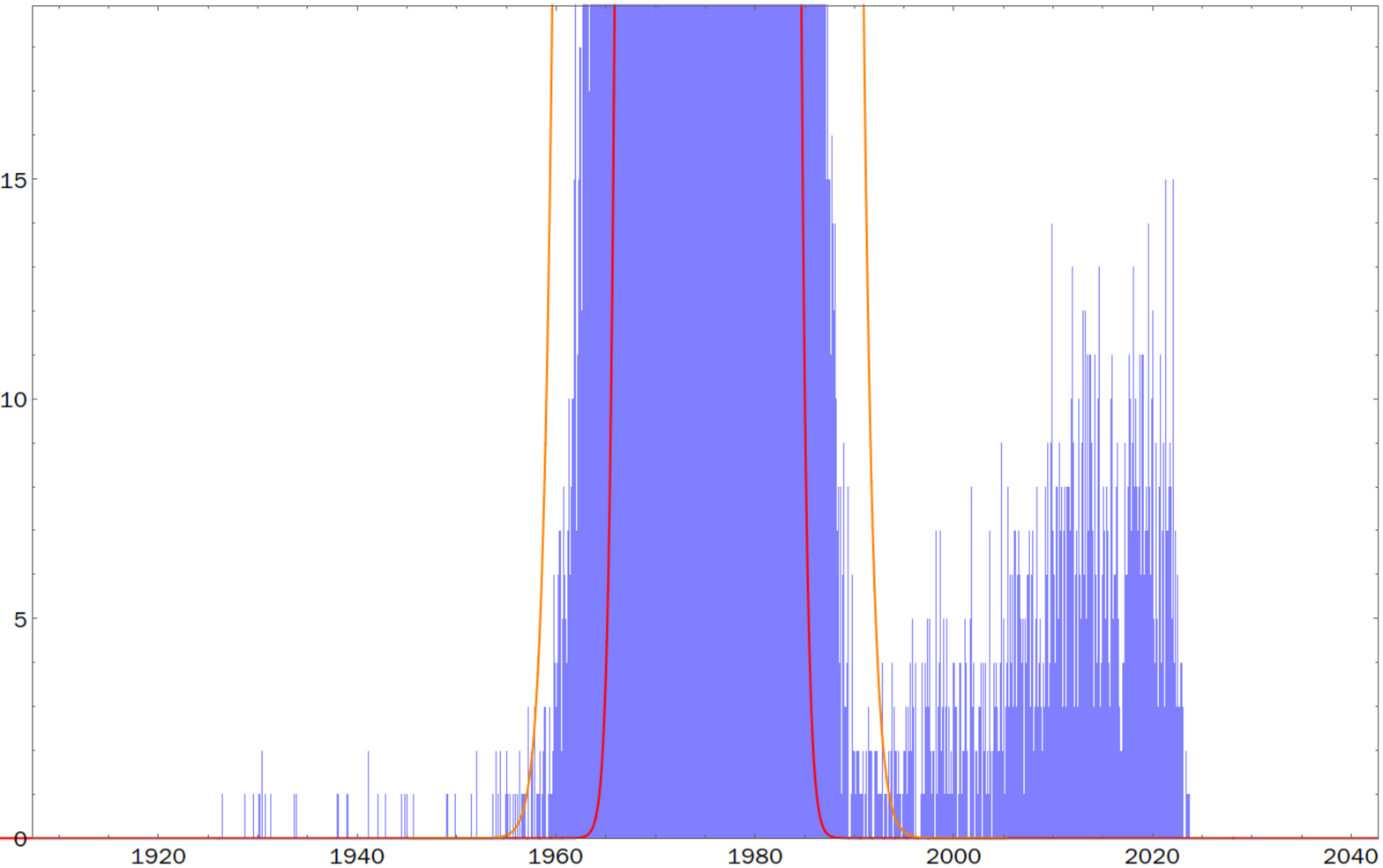}
  \end{minipage} \hfill
  \begin{minipage}[c]{0.35\textwidth}
    \caption{A zoom of $x$-axis region of Fig. \ref{fig:histmub} reveals extended but flat tails (note the scale of the $y$-axis).} \label{fig:zoom}
  \end{minipage}

\vspace{0.00mm}

  \begin{minipage}[c]{0.6\textwidth}
    \includegraphics[width=\textwidth]{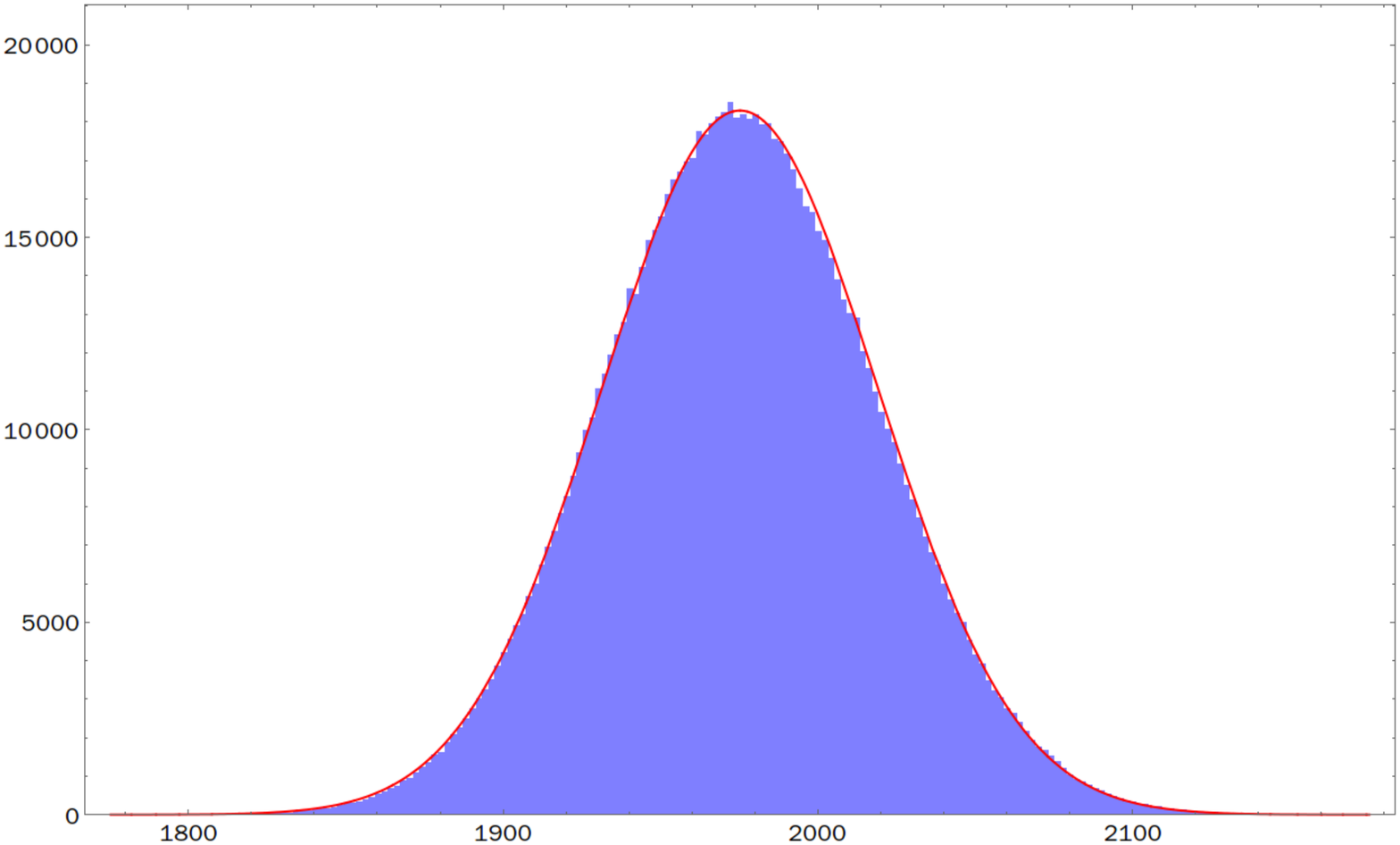}
  \end{minipage} \hfill
  \begin{minipage}[c]{0.35\textwidth}
    \caption{Histogram of estimations of $\text{Tr}\mathbb{O_F}$ with the Gaussian algorithm. The red line is a Gaussian distribution with the theoretical variance and the exact trace as mean. Same parameters as in Fig.~\ref{fig:histmub}.} \label{fig:gaussian}
  \end{minipage}
\end{figure}

Note that for such rather small lattices sizes it is possible to compute the matrix $O_F(0)O_F(m)^{-1}$ explicitly. Therefore one can test the
methods by comparing to the exact trace and exact variance and compute all $n(b-1)$ possible estimations of the MUBw method. For a realistic setup we use a configuration from an actual MC simulation with parameters~$N=8$,~$g=50$,~$L\,m=4$ and obtain an exact trace of $\approx1975.3$ and $V_\text{MUBw}\approx14.723$ and $V_\text{Gauss}\approx1920.1$ from the equations above. Also in this case the MUBw method yields a much smaller variance, which is  reflected in the histograms in Figs. \ref{fig:histmub} and \ref{fig:gaussian}. Since for the MUBw method the estimation can take on only a countable number of values Fig.~\ref{fig:histmub} contains all possible values, and the mean value reproduces the exact trace. In Fig.~\ref{fig:gaussian} there are one million trace estimations with the Gaussian method, whose histogram is very close to a Gaussian distribution with the exact trace as mean and $V_{Gauss}$ as variance (red curve), as expected. If we similarly compare the histogram in Fig.~\ref{fig:histmub} with a Gaussian distribution defined by the same mean and $V_{MUBw}$ as variance (orange curve), the peak obviously is much narrower than the exact variance suggests. Zooming in on the region close to the x-axis tails are revealed that are responsible for this mismatch. Fitting a Gaussian distribution to the central peak yields a variance of $\approx5.3$ (red curve). For illustration purpose the green curve is a best fit Cauchy distribution (where the height and position of the peak is fixed). The origin of the
tails in Fig.~\ref{fig:histmub} has not been studied, but in general will depend on the structure of the matrix. Contrary to the removal of the tail of zeros by excluding the standard basis (the difference between the original method and MUBw), there is no obvious way of removing them without introducing a bias. Without those tails one could safely estimate the trace with just one shot to a precision of $0.2\%$. But in the presence of these tails one should average over a number of estimates to exclude their dominance of the estimate.

\bigskip

We finally  comment on the speedups for using the MUBw estimator instead of the Gaussian one.  As explained above, the expectation value~\eqref{eq:TraceSimplification}  of the derivative of the null-cusp string effective action can be computed by Monte Carlo simulations (using RHMC). 
%Here we use a Rational Hybrid Monte Carlo algorithm~\cite{} generating a sample of field configurations on which the observables are evaluated. 
Estimators of the expectation value are then given by the sample mean and the stochastic error is proportional to the square root of the sample variance.

\begin{figure}
  \begin{minipage}[c]{0.72\textwidth}
    \includegraphics[width=\textwidth]{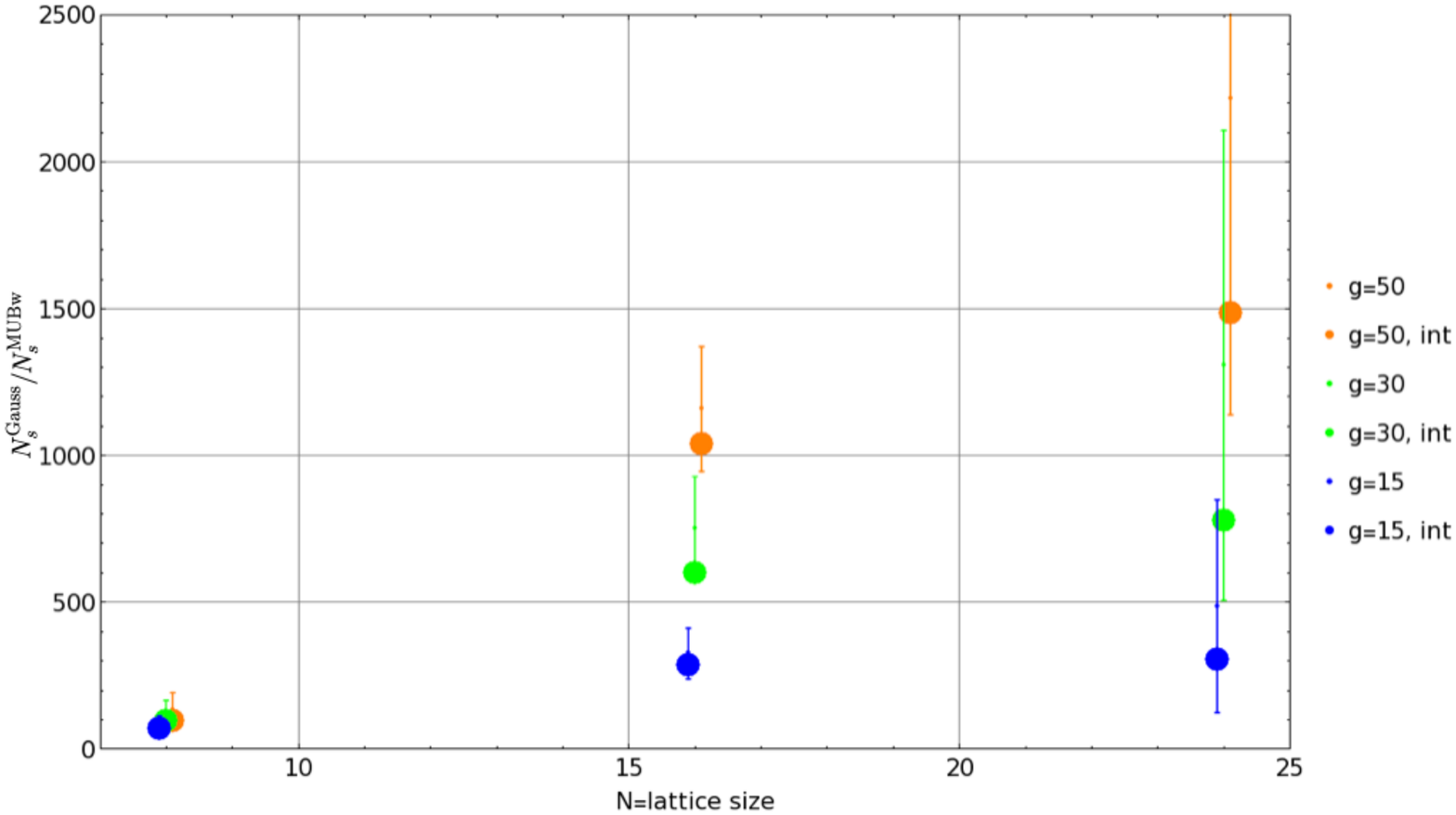}
  \end{minipage}\hfill
  \begin{minipage}[c]{0.25\textwidth}
    \caption{ Speedups of the MUBw algorithm compared with the Gaussian algorithm as a function of the lattice size $N$, for $N=8, 16, 24$. {Large circles correspond to an integer number of estimations, small dots are the results of equation $N_s=10\,V_\text{trace}/V_\text{RHMC}$.} } \label{fig:speedups}
  \end{minipage}
\end{figure}

To measure the observable of interest here, for each field configuration one needs to compute the trace~\eqref{trace}, which as explained is
most efficiently done by using an estimator based on matrix-vector products. Since stochastic estimators yield the trace with a stochastic error, such
a trace estimation embedded in the outer RHMC algorithm introduces additional fluctuations and thus an additional source of stochastic error. One way to deal with
this complication is to demand that the variance of the trace estimation $V_\text{trace}/N_s$~\footnote{Above, we denoted $V_\text{trace}$ simply as $V$.} is much smaller than the variance due to the field fluctuations $V_\text{RHMC}$, say a factor 10 smaller ($N_s=10\,V_\text{trace}/V_\text{RHMC}$, as in Fig.~\ref{fig:speedups}), so
that it contributes only a small fraction to the final error of the observable. For given variances this condition yields a value for the sample size $N_s$ of the trace estimation.

The numerical cost of the measurement of the trace contribution to the observable (cusp anomaly) is dominated by the trace estimation, more precisely by the multiplication of the inverse of $O_F$ with a vector. Therefore
the sample size $N_s$ is a direct measure of the numerical cost. Comparing different algorithms for the trace estimation leads to the ratio
\be\label{speedups}
N_s^\text{Gauss}/N_s^\text{MUBw}= V_\text{Gauss}/V_\text{MUBw}\,,
\ee
which defines the speedup for using MUBw instead of Gauss. This is a theoretical argument, where a speedup of 2 means that the algorithm finishes in half the time. In practice this depends also on the kind of implementation and the hardware used.
{In Fig.~\ref{fig:speedups} we see that the advantage of the MUBw algorithm grows with the lattice size and $g$. }

\section*{Acknowledgements}

We thank Johannes Weber for useful comments on the manuscript.  The research of VF received funding from the STFC grant ST/S005803/1, from the Einstein Foundation Berlin through an Einstein Junior Fellowship, and from the European Union Horizon 2020 research and innovation programme under the Marie Sklodowska-Curie grant agreement No 813942 ``Europlex''.  It was also supported in part by the Perimeter Institute for Theoretical Physics and the Simons Foundation through a Simons Emmy Noether Fellowship.  

  %%%%%%%%%%%%%%%%%%%%%%%%%%%%%%%%%%%%%
 %%%%%%%%%%%%%%%%%%%%%%%%%%%%%%%%%%%%%

\appendix

\section{The stringy setup}
\label{app:string}

In this Appendix we shortly describe the setup in which we test the algorithm.  The matrix differential operator we are interested in this note appears in the effective action for quantum fluctuations around the classical string solution~\cite{Giombi} which is dual, according to the AdS/CTF correspondence~\cite{MaldaWL}, to the cusp anomaly of $d=4$, $\mathcal{N}=4$ super Yang-Mills theory with gauge group $SU(N)$. The latter is a function $f(g)$ of the coupling $g=\sqrt{\lambda}/4\pi$ (where $\lambda$ is the t'Hooft coupling of the gauge theory) that governs the renormalization of a cusped, lightlike Wilson loop, and has an holographic definition via the path integral over quantum fluctuations around the string minimal surface ending on the cusped contour $C_\text{cusp}$ 
\be\label{Wcusp}
\langle W[C_\text{cusp}]\rangle \equiv Z_\text{cusp}=\int[D X][D \psi]\,e^{-S_\text{cusp}[X,\psi]}= e^{-\Gamma_\text{eff}}\equiv e^{- \frac{1}{8} f(g)\,m^2\,V_2} \,.
\ee
%Above, the first equivalence is one of the conjectures making the AdS/CFT correspondence, and the last one assumes. 
The path integral above is weighted by a (light-cone) gauge-fixed effective action $S_\text{cusp}$ of quantum fluctuation fields, defined on the two-dimensional string worldsheet parametrized by $(t,s)$, of bosonic ($X(t,s)$) and fermionic ($\psi(t,s)$) nature. 
Since the classical solution relevant here is homogenous, the worldsheet volume $V_2$ factorizes in front of $f(g)$ as in the last equivalence, where a massive parameter $m$ appears, related to the light-cone momentum of the string.  
The full form of the action, which is quartic in the fields, can be found in~\cite{Giombi} (or in~\cite{Bianchi:2016cyv,Bianchi:2019ygz}). After a Hubbard-Stratonovich linearization,  the fermionic sector becomes quadratic in the fields, thus making it possible to formally integrate out the fermion via a Pfaffian. One can then write %symbolically 
the action as 
%the sum of the bosonic lagrangian $\mathcal{L}_B$ (below we give its explicit, discretized version) and the fermionic one 
\begin{eqnarray}
\nonumber
S_\text{cusp}[X,\psi]&=&g\,\int dt \,ds\,\Big[\left|\partial_t x + \frac{m}{2} x \right|^2 + \frac{1}{z^4}\left|\partial_s x -\frac{m}{2} x \right|^2 + \left( \partial_t z^M+\frac{m}{2} z ^M\right)^2\\\label{Scusp}
&&
\qquad \qquad+\frac{1}{z^4}\left(\partial_s z^M - \frac{m}{2} z ^M\right)^2+\phi^2 + (\phi_I)^2
%\,\mathcal{L}_\text{B}[X]
+  \psi^{\text T} O_F[X,\phi, \phi_I] \psi \Big]\,.
\end{eqnarray}
 where  $g$ is the sigma-model effective coupling, which is with $m$ the other bare parameters of the model. The set of bosonic fields %~\footnote{In string theory, the coordinates of the spacetime where the string lives are actually fields...} 
 is given by the target space coordinates $x(t,s)$ and $x^*(t,s)$ (two complex bosonic $AdS_5$ coordinate fields orthogonal to the surface of the classical solution), $z^M$, with $\sqrt{z_M z^M}=z$ (further bosonic coordinates of $AdS_5 \times S^5$ in Poincar\'e patch), and the remaining bosonic fields $\phi, \phi_I$, $I=1,...,16$, which are  auxiliary fields arising from the Hubbard-Stratonovich transformation~\cite{Bianchi:2019ygz}. 
 
Above, $\psi\equiv({\theta}^i, { \theta}_i, {\eta}^i, {\eta}_i)$, $i=1,4$ is a vector collecting 8 complex Gra\ss mann variables $\theta$ and $\eta$ and their complex conjugates (for the notation and the different role of these variables see~\cite{Giombi, Bianchi:2019ygz}).  
% 
 %The vector $\psi$ in \eqref{Scuspquadratic} collects the 8 complex $\theta$ and $\eta$ in a formally ``redundant'' way which includes both the fields and their complex conjugates. Explicitating real and imaginary parts of $\theta,\eta$, it is easy to see that the fermionic contribution coming from this  $16\times 16$ complex operator $O_F$ is then the one of $16$ \emph{real} anti-commuting degrees of freedom.
 %
 The fermionic differential operator $O_F$ is the $16\times 16$ matrix~\eqref{OF}. 

\bigskip

In~\cite{Bianchi:2016cyv,Bianchi:2019ygz}, to obtain information about the function $f(g)$ the logaritmic derivative  of the partition function with respect to the coupling $g$ was considered, 
$
-g\frac{d\ln Z}{dg}=\langle S_\text{cusp}\rangle \equiv g V_2/8 \,f'(g)\,.
$
\\
%In this case one measures the vacuum expectation value of the action and obtains an estimation for $f'(g)$. 
A more direct definition of $f(g)$ uses the logarithmic derivative of the string partition function with respect to the massive parameter $m$ 
\be
-m \frac{\mathrm{d}\ln Z_\text{cusp}}{\mathrm{d} m} \equiv m \left\langle \frac{\mathrm{d}S_\text{cusp}}{\mathrm{d} m}\right\rangle%=-m \frac{\mathrm{d}}{\mathrm{d} m} \left( -\frac{1}{8} f(g) m^2 V_2 \right)
=\frac{m^2 V_2}{4}f(g) \,, 
\label{eq:mmfV}
\ee
where the derivative of the action is made of a bosonic and a fermionic contribution ($L_B$ is the bosonic lagrangian)
\begin{eqnarray} 
m \Big\langle  \frac{\mathrm{d}S_\text{cusp}}{\mathrm{d }  m}  \Big\rangle_{B,F}
 &=&m\,g\,\left \langle  \int dt ds \, \frac{\mathrm{d}}{\mathrm{d} m} L_B \right\rangle_{B,F}+m\,g\,\left\langle  \int dt ds\, \psi^{\text T} \frac{\mathrm{d}O_F} {\mathrm{d } m}  \psi  \right\rangle_{B,F}\,,
% &=&\left\langle m \int \dt \ds g \frac{\mathrm{d}}{\mathrm{d} m} \mathcal{L}_B \right\rangle_{F,B} + g m \left\langle \int \dt \ds \psi^{\text T} \frac{\mathrm{d}}{\mathrm{d} m} O_F \psi \right\rangle_{F,B}.
 \label{eq:mDmlnZ}
\end{eqnarray}
and $\langle ... \rangle_{B,F}= \frac{1}{Z} \int D  X D \psi (...) e^{-S[X,\psi]}$ is the standard vacuum expectation value, with $B,F$ indicating  path-integration over the fermionic and bosonic fields. 
In the second term above, the integration over fermions results in a term proportional to $O_F^{-1}\,\text{Pf}\, O_F$. Ignoring potential anomalies and working in a region of the parameter space where there is no sign change (see~\cite{Bianchi:2019ygz} for a detailed discussion~\footnote{We remark that the discretization of the adopted in~\cite{Bianchi:2019ygz} and used in this paper, despite breaking explicitly the $U(1)$ global symmetry of the model, was devised in such a way to obtain a real Pfaffian, albeit not definite positive. In a large region of the parameter space, which is the one adopted in the simulations of this paper, no sign change can be observed.}), the Pfaffian of the operator can be assumed to be positive definite and exponentiated via pseudofermions, 
${\rm Pf}\,O_F\equiv(\det O_F\,O^\dagger_F)^{\frac{1}{4}}= \int \!\!D\xi D\bar\xi\,e^{-\int dt ds\, \bar\xi(O_FO^\dagger_F)^{-\frac{1}{4}}\,\xi}$.  
Replacing now the continuum worldsheet with a square lattice of lattice spacing $a$, one arrives then to the symbolic expression
\begin{eqnarray} \label{dSdmintermediate}
m \Big\langle  \frac{\mathrm{d}S_\text{cusp}}{\mathrm{d }  m}  \Big\rangle_{B,F}
 &=&m\,g\,a^ 2\left \langle \sum \,  \frac{\mathrm{d}}{\mathrm{d} m} L_B \right\rangle_{B,P\!F}+\frac{m}{2}\, \left\langle \text{Tr}\,\Big(  \frac{\mathrm{d}O_F} {\mathrm{d } m}  \,O_F^{-1}\Big)  \right\rangle_{P\!F}\,,
% &=&\left\langle m \int \dt \ds g \frac{\mathrm{d}}{\mathrm{d} m} \mathcal{L}_B \right\rangle_{F,B} + g m \left\langle \int \dt \ds \psi^{\text T} \frac{\mathrm{d}}{\mathrm{d} m} O_F \psi \right\rangle_{F,B}.
\end{eqnarray}
where the sum replaces the integral over the continuum worldsheet  ($\sum\equiv\sum_{t,s}^{N_\text{t},N_\text{sp}}$, with $N_\text{sp}$ and $N_\text{t}$ the number of lattice points in the spatial and time directions, for simplicity we use a single parameter $N$ to denote the lattice points in spatial direction, the physical length is $L=a\cdot N$, the physical time in our simulations is $T=2L$). 
The second term above can be re-written using the fact that $O_F$ in~ \eqref{OF} is \emph{linear} in the parameter $m$. Then $m \frac{\mathrm{d}}{\mathrm{d} m} O_F(m)=O_F(m)-O_F(0)$, 
which can be used to express equation \eqref{dSdmintermediate} as
\be\!\!\!
m \Big\langle  \frac{\mathrm{d}S_\text{cusp}}{\mathrm{d }  m}  \Big\rangle_{B,F}
 =m\,g\,a^ 2\left \langle \sum \, \textstyle\frac{\mathrm{d}}{\mathrm{d} m} L_B \right\rangle_{B,P\!F}
-\frac{1}{2} \left\langle N_\text{lat}-\Tr \left(O_{F,Lat}(0)O_{F,Lat}(m)^{-1}\right)\right\rangle_{PF,B}\,,
\label{eq:TraceSimplification} 
\ee
where $N_\text{lat}=36\,N^2$ is the matrix size on the lattice. % being $N=N_s$ is the number of lattice points in spatial direction (the physical length is $L=a\cdot N$), and being the physical time $T=2L$.
%%%% Wilson discretization
Notice that the discretized setup of~\cite{Bianchi:2019ygz} adds to $O_F$ a Wilson-like lattice operator, in order to eliminate  fermionic doublers. In Fourier space, the resulting discretized  fermionic operator reads
\begin{eqnarray}\nonumber
\!\!\!\!\!\!\!\!\!\!\!\!\!\!\!\!
{\hat O_F}&\!\!=\!\!&\left(\begin{array}{cccc}
    \!\!\!\!  W_+      & \!\!\!\! - \mathring{p_0} \mathbb{1}      &\!\!\!\! (\mathring{p_1}-i\frac{m}{2} )\rho^M \frac{z^M}{z^3}  & \!\!\!\! 0 \\
 \!\!\!\!     - \mathring{p_0}\mathbb{1}  & \!\! -W_+^{\dagger}          &\!\!\!\! 0        & \!\!\!\! \rho_M^\dagger(\mathring{p_1}-i\frac{ m}{2})\frac{z^M}{z^3} \\
    \!\!\!\!  -( \mathring{p_1}+i\frac{m}{2} )\rho^M \frac{z^M}{z^3}   &\!\!\!\! 0          &\!\!\!\! 2\frac{{z}^{M}}{{z}^{4}}\rho^{M}\left(\partial_{s}{x}-m\frac{{x}}{2}\right) +W_-     &\!\!\!\! - \mathring{p_0} \mathbb{1}-A^T\\
   \!\!\!\!   0      &\!\!\!\! -\rho_M^\dagger( \mathring{p_1}+i\,\frac{m}{2})\frac{z^M}{z^3}   &\!\!\!\! - \mathring{p_0}\mathbb{1}+A    &\!\!\!\! -2\frac{{z}^{M}}{{z}^{4}}\rho_{M}^{\dagger}\left(\partial_{s}{x}^*-m\frac{{x}}{2}^*\right) -W_-^\dagger
          \end{array}\right)\\
          \label{OFWilson}
          \end{eqnarray}
          with
          \be\label{Wilsonshiftgen}
W_\pm = \frac{r}{2\,z^2}\,\big({\hat p}_0^2\pm i\,{\hat p}_1^2\big)\,\rho^M z_M\,,
\ee
$|r|=1$ and 
\be\label{pcirclephat}
\mathring{p}_\mu \equiv \frac{1}{a} \sin(p_\mu a)\,,\qquad\hat p_\mu\equiv \frac{2}{a} \sin\frac{p_\mu a}{2}\,.
\ee
%%%%%%%%

We also report here the relevant properties  of the fermionic operator $O_F$ useful for the evaluation of the trace in~\eqref{eq:TraceSimplification} above.
$O_F$ is antisymmetric, and satisfies the  constraint~\cite{Bianchi:2019ygz}
 \begin{equation}\label{gamma5prop}
O_F^\dagger=\Gamma_5\,O_F\, \Gamma_5
\end{equation}
where $\Gamma_5$ is the following unitary, antihermitian matrix
\be\label{Gamma5}
\Gamma_5=\left(\begin{array}{cccc}
			0 			&  \mathbb{1}			&0& 0 \\
			-\mathbb{1}	&0					&0				&0\\
			0&    0					&0				& \mathbb{1}\\
			0			&0	&-\mathbb{1}		&0
          \end{array}\right)\,,
\qquad   \Gamma_5^\dagger \Gamma_5=\mathbb{1}  \qquad \Gamma_5^\dagger=-\Gamma_5\,.
\ee
This ensures that  $\det O_F$ is \emph{real} and \emph{non-negative}, while the Pfaffian may in principle still suffer a sign problem. In this paper we do not explore the region of the parameter space (i.e. small values of $g_\text{LAT}$) for which this is true. 
One can check that the same properties also ensure that $\Tr \left(O_{F,Lat}(0)O_{F,Lat}(m)^{-1}\right)$ is real.

\bibliographystyle{nb}
\bibliography{Ref_strings_trace}

\end{document}